%% file: lblcpi_lifetime_prl.tex
\documentclass[aps,prl,twocolumn,superscriptaddress,showpacs,letterpaper]{revtex4}

\usepackage{graphicx}
\usepackage{amsmath, amssymb}

\newcommand{\Lb}{\ifmmode  \Lambda_b^0
                 \else
                           \mbox{$\Lambda_b^0$}
                 \fi
                }
\newcommand{\Lcpi}{\ifmmode  \Lambda_c^+\pi^-
                 \else
                           \mbox{$\Lambda_c^+\pi^-$}
                 \fi
                }
\newcommand{\Lc}{\ifmmode  \Lambda_c^+
                 \else
                           \mbox{$\Lambda_c^+$}
                 \fi
                }
\newcommand{\pkpi}{pK^-\pi^+}
\newcommand{\mkm}{\ifmmode \mu\rm{m}%
                 \else%
                           \mbox{$\mu$m}%
                 \fi%
                 }%
\newcommand{\gevc}{\ifmmode \rm{GeV}/c%
                 \else%
                           \mbox{GeV}$/c$%
                 \fi%
                 }%
\newcommand{\gevcc}{\ifmmode \rm{GeV}/c^2%
                 \else%
                           \mbox{GeV}$/c^2$%
                 \fi%
                 }%
\newcommand{\mevc}{\ifmmode \rm{MeV}/c%
                 \else%
                            \mbox{MeV}$/c$%
                 \fi%
                 }%
\newcommand{\mevcc}{\ifmmode \rm{MeV}/c^2%
                 \else%
                           \mbox{MeV}$/c^2$%
                 \fi%
                 }%
\newcommand{\ct}{\ifmmode ct%
                 \else%
                         \mbox{$ct$}%
                 \fi%
                 }%
\newcommand{\cterr}{\ifmmode \sigma_{ct}%
                 \else%
                         \mbox{$\sigma_{ct}$}%
                 \fi%
                 }%
\newcommand{\pt}{\ifmmode  p_{\rm T}%
                 \else%
                            \mbox{$p_{\rm{T}}$ }%
                 \fi%
                 }%
\newcommand{\ra}{\rightarrow}
\def\cdf2{CDF II}

\begin{document}

\title{Measurement of the \Lb Lifetime in 
       $\Lb \to \Lcpi$ Decays in $p\bar{p}$
       Collisions at $\sqrt{s} = 1.96$ TeV}

\input{October2009_Authors}


\begin{abstract}
We report a measurement of the lifetime of the \Lb baryon in decays to 
the $\Lcpi$ final state in a sample corresponding to $1.1$~fb$^{-1}$ 
collected in $p\bar{p}$ collisions at $\sqrt{s}=1.96$~TeV by the \cdf2 
detector at the Tevatron collider. Using a sample of about $3000$ fully 
reconstructed \Lb events we measure $\tau(\Lb) = 
1.401 \pm 0.046\mbox{ (stat)} \pm 0.035\mbox{ (syst)}~\mbox{ps}$ 
(corresponding to $c\tau(\Lb) = 
420.1 \pm 13.7\mbox{ (stat)} \pm 10.6\mbox{ (syst)}~\mu\mbox{m}$, 
where $c$ is the speed of light). The ratio of this result and the
world average $B^{0}$ lifetime yields $\tau(\Lb)/\tau(B^0) =
0.918 \pm 0.038$ (stat and syst), in good agreement with recent
theoretical predictions.
\end{abstract}

\maketitle

In the decays of beauty to charm hadrons the fundamental force underlying 
the decay of a \emph{b} quark to a \emph{c} quark is the weak interaction. 
However, the heavy \emph{b} quark is surrounded by a cloud of light quarks 
and gluons so the strong interaction must be taken into account. In the 
limit of an infinite mass of the \emph{b} quark, the heavy quark decouples 
from the light degrees of freedom. For a finite $m_b$ the decay rates can 
be computed as a series expanded in the small parameter $\Lambda_{QCD}/m_b$, 
where $m_b$ is the mass of the \emph{b} quark and $\Lambda_{QCD}$ is the 
energy scale of the QCD interactions within the hadron. This is known as 
the heavy-quark expansion (HQE)~\cite{Bellini:1996ra}.  The application of 
HQE to the decays of the $\Lb$ baryon (\emph{udb}) and the beauty mesons 
$(B^{0}, \overline{b}d; B^{+}, \overline{b}u)$ does not result in an 
identical series. For example, in the $(\Lambda_{QCD}/m_b)^{3}$ term 
\emph{W}-boson exchange contributions are quite different~\cite{Bigi:1999dq}, 
leading to a prediction of $\tau(\Lb)/\tau(B^0) \neq 1$. Experimental
studies of beauty hadron lifetimes therefore help us to test the theoretical
understanding of the HQE series, and consequently the underlying QCD physics.

Over the past five years theoretical predictions of the lifetime ratio
$\tau(\Lb)/\tau(B^0)$ have not agreed with experimental values. In 
2004 an HQE calculation including $\mathcal{O}(1/m_{b}^{4})$ effects resulted 
in $\tau(\Lb)/\tau(B^0)=0.86 \pm 0.05$~\cite{Gabbiani:2004tp}. This 
was in good agreement with the 2006 experimental world average of 
$0.804 \pm 0.049$~\cite{bib:pdg2006}. In 2006 the CDF collaboration reported 
a measurement~\cite{Abulencia:2006dr} of the \Lb lifetime in the 
$\Lb \to J/\psi \Lambda^0$ channel such that $\tau(\Lb)/\tau(B^0)$
differed by $+2\sigma$ from the 2006 world average~\cite{bib:pdg2006}, was 
significantly higher than the 2004 HQE calculation~\cite{Gabbiani:2004tp}, 
but was compatible with earlier HQE predictions~\cite{bib:early_pred}.  A 
more recent measurement by the D\O\ collaboration~\cite{Abazov:2007sf} in 
the same channel leads to a value of $\tau(\Lb)/\tau(B^0)$ which is 
compatible with both the 2006 world average~\cite{bib:pdg2006} and the CDF
value~\cite{Abulencia:2006dr}.

In this paper we present the first measurement of the \Lb lifetime in a 
fully hadronic final state. The data sample is produced in $p\bar{p}$ 
collisions at $\sqrt{s} = 1.96$~TeV at the Tevatron and corresponds to an
integrated luminosity of $1.1~\mathrm{fb^{-1}}$. We reconstruct \Lb in the 
$\Lb\to\Lcpi$ decay channel where \Lc subsequently decays as 
$\Lc \ra \pkpi$. Throughout the paper, reference to  a specific charge 
state also implies the charge conjugate state. 

The components of the \cdf2 detector~\cite{Abulencia:2005ix} most relevant 
for this analysis are the tracking system and the displaced vertex trigger 
system. The tracking system lies within a uniform axial magnetic field of
$1.4$~T.  The inner tracking volume is instrumented with either 6 or 7
layers of double-sided silicon microstrip detectors up to a radius of
$28$~cm from the beamline~\cite{Sill:2000zz}. These surround a layer of
single-sided silicon mounted directly on the beam pipe at a radius of
$1.5$~cm~\cite{Hill:2004qb}. This system provides an excellent resolution 
(about $40~\mkm$) on the impact parameter $(d_0)$, which is defined as the 
distance of closest approach of the charged particle to the $p\bar{p}$ 
interaction point in the plane transverse to the beam direction. The $d_0$ 
resolution of $40~\mkm$ includes an approximate $30~\mkm$ contribution from 
the uncertainty of the interaction point in the transverse plane (added in 
quadrature). The outer tracking volume contains an open-cell drift chamber 
(COT) up to the radius of $137$~cm\cite{Affolder:2003ep}.

\cdf2 employs a three-level trigger system. The extremely fast tracker 
(XFT)~\cite{Thomson:2002xp} at the first level groups COT hits into tracks 
in the transverse plane. At the second level, the silicon vertex trigger 
(SVT)~\cite{Adelman:2007zz} adds silicon hits to the tracks found by the 
XFT, improving the resolution of the track position and thus allowing 
selection based on the transverse displacement from the beam line that is 
measured in real time. The displaced vertex trigger~\cite{Abulencia:2006df}
requires two charged particles with momentum transverse to the beam
direction (\pt) greater than $2$~\gevc, and with impact parameters
in the range $0.12 < |d_0| < 1~\mbox{mm}$.  The intersection point of
the two particle trajectories must have a transverse displacement 
($L_{xy}$) from the interaction point of at least $200~\mkm$. The pair must 
also have a scalar sum $p_T(1)+p_T(2) > 5.5$~\gevc.  This trigger 
configuration based on a pair of tracks is called the two-track trigger 
(TTT) and is the basis for the collection of many fully hadronic bottom and 
charm decays at CDF.

We reconstruct a \Lb candidate via its decay to \Lc $\pi^-$, where the \Lc 
further decays to a $\pkpi$ final state.  All four tracks are required 
to have a sufficient number of hits in the tracking detectors for 
high-quality position measurement. Several requirements are imposed to 
suppress background in the reconstructed sample which are optimized using 
simulated signal and data background samples~\cite{sigma:2007rw}. Each 
particle must have $|d_0| < 1000~\mkm$. We construct \Lc candidates by 
combining three tracks assuming the ($\pkpi$) hypothesis. The $p$ 
candidate and the $\pi^-$ are required to have $\pt > 2.0$~\gevc. The proton 
\pt must exceed the \pt of the $\pi^+$ from the \Lc, which has the same 
charge. This prevents the same pair of tracks being considered both as 
$(p,\pi^{+})$ and as $(\pi^{+},p)$. The three tracks from the \Lc candidate 
are first constrained to a common vertex in  a kinematic fit. Next we add a 
track and construct \Lb candidates through a further kinematic fit, which 
intersects the fourth track with the \Lc candidate trajectory. The mass of 
the \Lc candidate is constrained to the world average \Lc mass 
($2.286$~\gevcc)~\cite{pdg09}. This second kinematic fit allows us to 
calculate \ct = $L_{xy} c M/\pt$ and its uncertainty, \cterr, where $c$ 
is the speed of light, and $t$ and $M$ are the proper decay time and 
measured mass of the \Lb, respectively.

We apply additional selection requirements in order to suppress background.
The requirements on $\ct(\Lb) > 250~\mkm$, its significance
$\ct(\Lb)/\cterr>10$, and $\left |d_{0}(\Lb)\right | < 80~\mkm$
primarily suppress the background arising from random combinations of tracks, 
many of which originate from the primary interaction point (combinatorial 
background). Another important source of background is the decay of $B$
mesons with misidentified decay products.  Decays like $\bar{B}^0 \to
D^+\pi^-$ are especially insidious since they are abundant (compared to 
$\Lb \to \Lcpi$ decays) and $D^+ \to K^-\pi^+\pi^+$ decays can easily mimic 
the $\Lc\to K^- p \pi^+$ signature.  These backgrounds are suppressed by 
selecting a narrow region of the invariant mass spectrum of the $\Lc \to 
\pkpi$ candidate: $\left | m(\pkpi)-m(\Lc)_{PDG} \right | <16$~\mevcc. 
Further $D^+$ candidates are removed by a requirement on the $ct$ of \Lc 
candidates with respect to the \Lb vertex, since the \Lc candidates are 
usually much shorter lived than the $D^+$ candidates. We require 
$-70 < \ct(\Lc \textrm{w.r.t.} \Lb) < 200~\mkm$. Lastly, the TTT criteria 
are confirmed using the reconstructed candidate tracks.

The lifetime of the \Lb baryon is determined from two sequential maximum 
likelihood fits. The first is a fit to the invariant mass of $\Lcpi$ 
candidates and is used to establish the composition of the sample.  This 
gives the normalization of each of the fit components for both the whole 
domain of $4.82 < m(\Lb) < 7.0$~\gevcc, as well as the signal region 
($5.565 < m(\Lb) < 5.670$~\gevcc).  The second fit is an unbinned maximum 
likelihood fit of \ct and \cterr in the signal region to extract the 
\Lb lifetime with the normalizations of each component fixed.

\begin{figure}[htbp]
\centering
\includegraphics[width=0.48\textwidth]{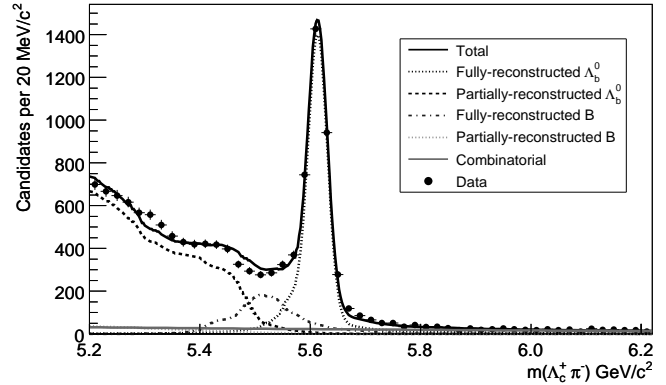}
\caption{The distribution of the invariant mass of $\Lb \ra \Lcpi$ 
         candidates (points) with the fit overlaid (solid black line).}
\label{fig:lb_mass}
\end{figure}

The invariant mass distribution of \Lcpi candidates is shown in
Fig.~\ref{fig:lb_mass} with the fit projection overlaid.  Small deviations
of the model from data below the $\Lb$ mass do not affect the lifetime as 
they occur outside the signal region. The \Lcpi mass distribution is 
described by several components: the $\Lb\to\Lcpi$ signal, a combinatorial 
background, partially and fully reconstructed $B$ mesons that pass the \Lcpi
selection criteria, partially reconstructed \Lb decays, and fully 
reconstructed \Lb decays other than \Lcpi ({\it e.g.} $\Lb \to \Lc K^-$). 
The combinatorial background is modeled with an exponentially decreasing 
function of \Lcpi mass.  All other components are represented in the 
f\mbox{}it by f\mbox{}ixed shapes derived from Monte Carlo (MC)
simulations~\cite{inclusive-b-samples} whose relative contributions are 
constrained using data when possible. Significant differences between fit 
and data are only observed outside the signal region. The mass fit has 
$2905 \pm 58$ $\Lb \ra \Lcpi$ signal events, $252 \pm 46$ other fully 
reconstructed \Lb candidates (which are also used to determine the \Lb 
lifetime), and $11\%$ background in the signal region.

Due to the trigger requirements on the track $d_0$ and track-pair $L_{xy}$, 
the observed \Lb \ct distribution is not a simple exponential. 
Consequently, an efficiency ($\epsilon(ct)$) must be included to model the 
acceptance of the trigger and offline selection. The largest corrections are 
due to the $d_0$ requirements of the TTT. The two-dimensional $\ct-\cterr$ 
probability density function (pdf) for the signal and other fully 
reconstructed \Lb components is given by
\begin{equation}
 P(ct,c\tau,\cterr;S_{1,2}) =
        P(ct|c\tau,\cterr,S_{1,2}) \cdot P(\cterr) \cdot
        \epsilon(ct).
  \label{eq:ctwitheff-with-Sct}
\end{equation}
where $S_{1,2}$ are the two \cterr scale factors obtained from a
two-Gaussian modeling of the resolution function in MC (one for each
Gaussian). The scale factor is necessary because the kinematic fitter
underestimates the uncertainty on the \ct (the same scale factor is used 
for signal and background). $P(ct|\cterr,S_{1,2})$ is a one-dimensional 
conditional pdf for observing this value of \ct given the true \Lb 
lifetime ($\tau$), \cterr, and $S_{1,2}$. For the fully reconstructed \Lb 
components this pdf is a decreasing exponential convoluted with the sum of 
the two resolution Gaussians. $P(\cterr)$ is the pdf for observing 
$\cterr$ and is obtained from the sideband subtracted data distribution, 
where the sideband is defined as $5.8 < m(\Lb) < 7.0$~\gevcc. For each 
background component Eq.~\ref{eq:ctwitheff-with-Sct} is modified in a 
suitable way, apart from the partially reconstructed $B$ mesons, which do 
not populate the signal region and are therefore not included in the 
lifetime fit.

\begin{figure}[htbp]
\centering
\includegraphics[width=0.48\textwidth]{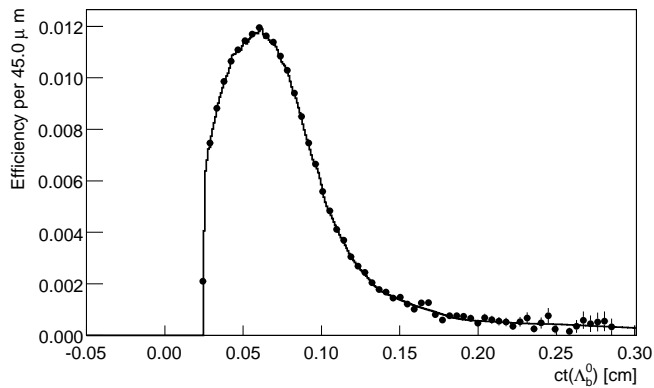}
\caption{The smoothed \Lb TTT efficiency histogram (line) superimposed on 
         the unsmoothed histogram (points)}
\label{fig:SVT_Eff}
\end{figure}

A sample of simulated signal events is used to extract  $\epsilon(ct)$. This 
sample consists of single $b$ hadrons generated with a $\pt$ spectrum 
extracted from the data sample and decayed with EvtGen~\cite{Lange:2001uf}. 
This MC sample is further reweighted in order to match the data in a number 
of relevant variables: the choice of `trigger tracks' (the pair of final 
state particles which cause the TTT to fire), the proton production angle in 
\Lc\ rest frame which is sensitive to \Lb\ polarization, and the 
contributions of the \Lc\ Dalitz components~\cite{Dalitz:1953cp}. The TTT 
efficiency function is represented by a histogram calculated as
\(\epsilon(ct) = h(ct) / 
        \sum_{i} exp(\ct,c\tau^{MC})
                \otimes R(S_{1,2},\cterr^{i})\).
The numerator is a smoothed histogram of the \ct for all MC events that 
pass the trigger and analysis selection criteria. Each bin of the 
denominator is calculated by summing the analytical \ct distribution at the 
\ct bin center over all events (indexed by $i$) that pass the criteria 
required to fill the numerator. The analytical \ct distribution is an 
exponential convoluted with the resolution function $R$. 
Fig.~\ref{fig:SVT_Eff} shows the resulting finely binned TTT efficiency 
histogram used for the \Lb signal components. Exactly the same procedure was 
used to derive an efficiency histogram for the fully reconstructed $B$ meson 
background.

Our approach assumes that the simulation of trigger and detector can be used 
to derive $\epsilon(ct)$. This assumption can be validated in data using 
$J/\psi \to \mu^+\mu^-$ decays collected by the dimuon trigger which does not 
bias their lifetime. The observed four momenta of $J/\psi \to \mu^+\mu^-$ 
decays were also used as the input for a simulated sample of 
$J/\psi \to \mu^+\mu^-$ decays, subsequently fed to the TTT and detector 
simulation (data-seeded MC). Comparing the number of real 
$J/\psi \to \mu^+\mu^-$ decays that pass the TTT with the number of 
data-seeded simulation decays that pass the TTT simulation gives a direct 
check of the reliability of the simulation. For both real and simulated 
$J/\psi$ decays we compute the TTT efficiency as a function of $L_{xy}$ and 
form their ratio $R_{\epsilon}(L_{xy})$. The deviation of the slope of 
$R_{\epsilon}(L_{xy})$ from $0$ is a measure of the quality of modeling of 
TTT in the simulation.  Observed $R_{\epsilon}(L_{xy})$ is incompatible with 
a null slope at the $3-4\sigma$ level; we treat this discrepancy as a source 
of systematic uncertainty.

\begin{figure}[htbp]
\centering
\includegraphics[width=0.48\textwidth]{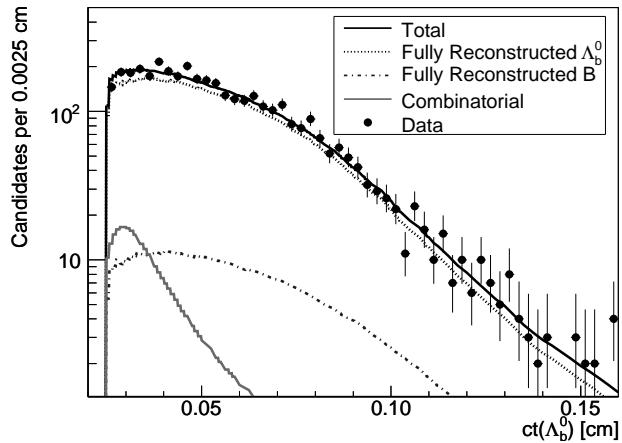}
\caption{The distribution of the \ct of $\Lb \ra \Lcpi$ candidates (points)
         with the fit projection overlaid (solid black line). The partially 
         reconstructed \Lb fit components only have a $1\%$ contribution to 
         the signal region and are not shown.}
\label{fig:lifeFit}
\end{figure}

We perform an unbinned maximum likelihood fit for $c\tau(\Lb)$ to the data 
that yields $c\tau(\Lb) = 420.1 \pm 13.7~\mkm$ (stat). The total likelihood 
is:
\( \mathcal{L} = \prod_i \sum_j N^{\rm sig}_{j} 
\cdot P_{j}(ct_{i},\sigma_{ct\;i};S_{1,2})\),
where the subscript $i$ runs over events, and the subscript $j$ runs over 
classes of event: the fully reconstructed signal and background fit 
components, the combinatorial background, and the partially reconstructed 
\Lb decays. $P_{j}(ct_{i},\sigma_{ct\;i};S_{1,2})$ is a two-dimensional pdf
of the form given in Eq.~\ref{eq:ctwitheff-with-Sct}, $N^{\rm sig}_{j}$ is 
the number of events of this class occurring in the signal region. The 
resulting likelihood projected onto the \ct-axis is shown in 
Fig.~\ref{fig:lifeFit}. We fit the data for the \Lb 
lifetime after all procedures are established. The fit probability is 
estimated to be $37\%$ using an unbinned Kolmogorov-Smirnov test.

For each source of systematic uncertainty, we generate sets of events for 
about $500$ pseudo experiments from a modified pdf and fit with both the 
standard and modified fits. The mean of the distribution of the difference 
between fit results obtained with the standard and modified pdf is used as 
the systematic uncertainty. We consider the systematic uncertainties in two 
groups based on whether they affect the TTT efficiency or not.

In the first group, the systematic uncertainty due to the alignment of the 
silicon detector is quoted from a previous study~\cite{Abulencia:2006dr} 
($2.0~\mkm$) where internal silicon sensor deformations and global 
misalignments of the silicon detector relative to the outer tracking volume 
are taken into account. The uncertainty due to the background component 
normalizations was taken into account by varying them according to their 
uncertainties derived from the mass fit ($1.0~\mkm$). 

In the second group of uncertainties, where the TTT efficiency is directly 
affected, the leading source is due to the slope of $R_{\epsilon}(L_{xy})$ 
($8.6~\mkm$). The uncertainty due to the \Lc Dalitz structure is evaluated 
by varying the relative contributions of each Dalitz component according to 
the world average uncertainty~\cite{pdg09} ($3.7~\mkm$). The effect of an 
uncertainty in the combinatorial background \ct template is computed by
modifying it to a smoothed version of the actual upper sideband \ct
distribution ($2.9~\mkm$). The uncertainty due to the particle identity of 
the tracks which fired the trigger is evaluated by varying the relative
contributions of different trigger-track combinations in the MC ($2.0~\mkm$). 
The uncertainty due to the \Lb polarization is obtained by varying the slope 
of the MC reweighting factor by one-sigma from a straight line fit for the 
proton production angle in the \Lc rest frame ($1.4~\mkm$). The uncertainty 
due to the transverse position of the $p\bar{p}$ primary interaction point 
is computed by dividing the MC into independent subsamples representing the 
extreme variations of the primary interaction point ($1.2~\mkm$). The 
uncertainty due to the TTT efficiency used for the $B^{0}$ background is 
evaluated by tightening the mass cut on the $D^{+}\pi^{-}$ candidate in the 
underlying $B^{0}$ MC reconstruction ($1.0~\mkm$). The uncertainty due to
the lifetime assumed for the $B^{0}$ background is obtained by varying this 
lifetime according to the world average uncertainty~\cite{pdg09} 
($1.0~\mkm$). The uncertainty due to a correlation between the $d_0$ 
requirements in the SVT and reconstruction levels is estimated by smearing 
the latter in the signal MC by an amount extracted by comparing their 
difference distributions between data and MC ($1.0~\mkm$). The total 
systematic uncertainty is computed by adding all the contributions in 
quadrature, which is $10.6~\mkm$.

Numerous cross-checks were performed. We used our procedure to measure the 
$B^0$ lifetime in the $B^0 \ra D^{*-}\pi^{+}$ and $B^0 \ra D^{-}\pi^{+}$ 
decay modes and the $B^+$ lifetime in the $B^+ \ra \overline{D}^{0} \pi^{+}$ 
mode. These $B^0$ and $B^+$ lifetime measurements are statistically 
consistent with the world averages~\cite{pdg09}. We checked the effect of 
uncertainties in the mass template shapes, the \pt spectrum of the \Lb, 
the effect of assuming different \Lb lifetimes in the MC, the scale factor 
applied to the \cterr, the \Lc lifetime, and the model of the
uncertainty of the transverse position of the $p\bar{p}$ primary interaction 
point. We also checked the effect of the uncertainty in the shape of 
\cterr and used a large signal MC sample to verify that the fitter 
itself does not introduce a bias in the measured lifetime.

In summary, using a sample of $2905 \pm 58$ fully reconstructed  $\Lb \to 
\Lcpi$ decays we measure the lifetime of the \Lb baryon to be $\tau(\Lb) 
= 1.401 \pm 0.046\mbox{ (stat)} \pm 0.035\mbox{ (syst)}~\mbox{ps}$ 
(corresponding to $c\tau(\Lb) = 420.1 \pm 13.7\mbox{ (stat)} \pm 10.6\mbox{ 
(syst)}~\mu\mbox{m}$ where $c$ is the speed of light). This is the single 
most precise measurement of the \Lb lifetime. 

Using the current world average for the $B^0$ lifetime~\cite{pdg09}, we 
obtain $\tau(\Lb)/\tau(B^0) = 0.918 \pm 0.038$ (stat + syst). There 
is good agreement between our result and the current world average of 
$\tau(\Lb)/\tau(B^0) = 0.99 \pm 0.10$~\cite{pdg09}, and between our 
result and the previous CDF result\cite{Abulencia:2006dr}. This measurement 
is also compatible with the current HQE value~\cite{Gabbiani:2004tp} of 
$\tau(\Lb)/\tau(B^0) = 0.86 \pm 0.05$, thus supporting the HQE 
picture of weak decays of heavy baryons.

We thank the Fermilab staff and the technical staffs of the participating 
institutions for their vital contributions. This work was supported by the 
U.S. Department of Energy and National Science Foundation; the Italian 
Istituto Nazionale di Fisica Nucleare; the Ministry of Education, Culture, 
Sports, Science and Technology of Japan; the Natural Sciences and Engineering 
Research Council of Canada; the National Science Council of the Republic of 
China; the Swiss National Science Foundation; the A.P. Sloan Foundation; the 
Bundesministerium f\"ur Bildung und Forschung, Germany; the World Class 
University Program, the National Research Foundation of Korea; the Science 
and Technology Facilities Council and the Royal Society, UK; the Institut 
National de Physique Nucleaire et Physique des Particules/CNRS; the Russian 
Foundation for Basic Research; the Ministerio de Ciencia e Innovaci\'{o}n, 
and Programa Consolider-Ingenio 2010, Spain; the Slovak R\&D Agency; and the 
Academy of Finland.


\end{document}

%% file: October2009_Authors.tex
\affiliation{Institute of Physics, Academia Sinica, Taipei, Taiwan 11529, Republic of China} 
\affiliation{Argonne National Laboratory, Argonne, Illinois 60439} 
\affiliation{University of Athens, 157 71 Athens, Greece} 
\affiliation{Institut de Fisica d'Altes Energies, Universitat Autonoma de Barcelona, E-08193, Bellaterra (Barcelona), Spain} 
\affiliation{Baylor University, Waco, Texas  76798} 
\affiliation{Istituto Nazionale di Fisica Nucleare Bologna, $^{cc}$University of Bologna, I-40127 Bologna, Italy} 
\affiliation{Brandeis University, Waltham, Massachusetts 02254} 
\affiliation{University of California, Davis, Davis, California  95616} 
\affiliation{University of California, Los Angeles, Los Angeles, California  90024} 
\affiliation{University of California, San Diego, La Jolla, California  92093} 
\affiliation{University of California, Santa Barbara, Santa Barbara, California 93106} 
\affiliation{Instituto de Fisica de Cantabria, CSIC-University of Cantabria, 39005 Santander, Spain} 
\affiliation{Carnegie Mellon University, Pittsburgh, PA  15213} 
\affiliation{Enrico Fermi Institute, University of Chicago, Chicago, Illinois 60637}
\affiliation{Comenius University, 842 48 Bratislava, Slovakia; Institute of Experimental Physics, 040 01 Kosice, Slovakia} 
\affiliation{Joint Institute for Nuclear Research, RU-141980 Dubna, Russia} 
\affiliation{Duke University, Durham, North Carolina  27708} 
\affiliation{Fermi National Accelerator Laboratory, Batavia, Illinois 60510} 
\affiliation{University of Florida, Gainesville, Florida  32611} 
\affiliation{Laboratori Nazionali di Frascati, Istituto Nazionale di Fisica Nucleare, I-00044 Frascati, Italy} 
\affiliation{University of Geneva, CH-1211 Geneva 4, Switzerland} 
\affiliation{Glasgow University, Glasgow G12 8QQ, United Kingdom} 
\affiliation{Harvard University, Cambridge, Massachusetts 02138} 
\affiliation{Division of High Energy Physics, Department of Physics, University of Helsinki and Helsinki Institute of Physics, FIN-00014, Helsinki, Finland} 
\affiliation{University of Illinois, Urbana, Illinois 61801} 
\affiliation{The Johns Hopkins University, Baltimore, Maryland 21218} 
\affiliation{Institut f\"{u}r Experimentelle Kernphysik, Karlsruhe Institute of Technology, D-76131 Karlsruhe, Germany} 
\affiliation{Center for High Energy Physics: Kyungpook National University, Daegu 702-701, Korea; Seoul National University, Seoul 151-742, Korea; Sungkyunkwan University, Suwon 440-746, Korea; Korea Institute of Science and Technology Information, Daejeon 305-806, Korea; Chonnam National University, Gwangju 500-757, Korea; Chonbuk National University, Jeonju 561-756, Korea} 
\affiliation{Ernest Orlando Lawrence Berkeley National Laboratory, Berkeley, California 94720} 
\affiliation{University of Liverpool, Liverpool L69 7ZE, United Kingdom} 
\affiliation{University College London, London WC1E 6BT, United Kingdom} 
\affiliation{Centro de Investigaciones Energeticas Medioambientales y Tecnologicas, E-28040 Madrid, Spain} 
\affiliation{Massachusetts Institute of Technology, Cambridge, Massachusetts  02139} 
\affiliation{Institute of Particle Physics: McGill University, Montr\'{e}al, Qu\'{e}bec, Canada H3A~2T8; Simon Fraser University, Burnaby, British Columbia, Canada V5A~1S6; University of Toronto, Toronto, Ontario, Canada M5S~1A7; and TRIUMF, Vancouver, British Columbia, Canada V6T~2A3} 
\affiliation{University of Michigan, Ann Arbor, Michigan 48109} 
\affiliation{Michigan State University, East Lansing, Michigan  48824}
\affiliation{Institution for Theoretical and Experimental Physics, ITEP, Moscow 117259, Russia} 
\affiliation{University of New Mexico, Albuquerque, New Mexico 87131} 
\affiliation{Northwestern University, Evanston, Illinois  60208} 
\affiliation{The Ohio State University, Columbus, Ohio  43210} 
\affiliation{Okayama University, Okayama 700-8530, Japan} 
\affiliation{Osaka City University, Osaka 588, Japan} 
\affiliation{University of Oxford, Oxford OX1 3RH, United Kingdom} 
\affiliation{Istituto Nazionale di Fisica Nucleare, Sezione di Padova-Trento, $^{dd}$University of Padova, I-35131 Padova, Italy} 
\affiliation{LPNHE, Universite Pierre et Marie Curie/IN2P3-CNRS, UMR7585, Paris, F-75252 France} 
\affiliation{University of Pennsylvania, Philadelphia, Pennsylvania 19104}
\affiliation{Istituto Nazionale di Fisica Nucleare Pisa, $^{ee}$University of Pisa, $^{ff}$University of Siena and $^{gg}$Scuola Normale Superiore, I-56127 Pisa, Italy} 
\affiliation{University of Pittsburgh, Pittsburgh, Pennsylvania 15260} 
\affiliation{Purdue University, West Lafayette, Indiana 47907} 
\affiliation{University of Rochester, Rochester, New York 14627} 
\affiliation{The Rockefeller University, New York, New York 10021} 
\affiliation{Istituto Nazionale di Fisica Nucleare, Sezione di Roma 1, $^{hh}$Sapienza Universit\`{a} di Roma, I-00185 Roma, Italy} 

\affiliation{Rutgers University, Piscataway, New Jersey 08855} 
\affiliation{Texas A\&M University, College Station, Texas 77843} 
\affiliation{Istituto Nazionale di Fisica Nucleare Trieste/Udine, I-34100 Trieste, $^{ii}$University of Trieste/Udine, I-33100 Udine, Italy} 
\affiliation{University of Tsukuba, Tsukuba, Ibaraki 305, Japan} 
\affiliation{Tufts University, Medford, Massachusetts 02155} 
\affiliation{Waseda University, Tokyo 169, Japan} 
\affiliation{Wayne State University, Detroit, Michigan  48201} 
\affiliation{University of Wisconsin, Madison, Wisconsin 53706} 
\affiliation{Yale University, New Haven, Connecticut 06520} 
\author{T.~Aaltonen}
\affiliation{Division of High Energy Physics, Department of Physics, University of Helsinki and Helsinki Institute of Physics, FIN-00014, Helsinki, Finland}
\author{J.~Adelman}
\affiliation{Enrico Fermi Institute, University of Chicago, Chicago, Illinois 60637}
\author{B.~\'{A}lvarez~Gonz\'{a}lez$^v$}
\affiliation{Instituto de Fisica de Cantabria, CSIC-University of Cantabria, 39005 Santander, Spain}
\author{S.~Amerio$^{dd}$}
\affiliation{Istituto Nazionale di Fisica Nucleare, Sezione di Padova-Trento, $^{dd}$University of Padova, I-35131 Padova, Italy} 

\author{D.~Amidei}
\affiliation{University of Michigan, Ann Arbor, Michigan 48109}
\author{A.~Anastassov}
\affiliation{Northwestern University, Evanston, Illinois  60208}
\author{A.~Annovi}
\affiliation{Laboratori Nazionali di Frascati, Istituto Nazionale di Fisica Nucleare, I-00044 Frascati, Italy}
\author{J.~Antos}
\affiliation{Comenius University, 842 48 Bratislava, Slovakia; Institute of Experimental Physics, 040 01 Kosice, Slovakia}
\author{G.~Apollinari}
\affiliation{Fermi National Accelerator Laboratory, Batavia, Illinois 60510}
\author{A.~Apresyan}
\affiliation{Purdue University, West Lafayette, Indiana 47907}
% ...

\collaboration{CDF Collaboration\footnote{With visitors from $^a$University of Massachusetts Amherst, Amherst, Massachusetts 01003,
$^b$Universiteit Antwerpen, B-2610 Antwerp, Belgium, 
$^c$University of Bristol, Bristol BS8 1TL, United Kingdom,
$^d$Chinese Academy of Sciences, Beijing 100864, China, 
$^e$Istituto Nazionale di Fisica Nucleare, Sezione di Cagliari, 09042 Monserrato (Cagliari), Italy,
$^f$University of California Irvine, Irvine, CA  92697, 
$^g$University of California Santa Cruz, Santa Cruz, CA  95064, 
$^h$Cornell University, Ithaca, NY  14853, 
$^i$University of Cyprus, Nicosia CY-1678, Cyprus, 
$^j$University College Dublin, Dublin 4, Ireland,
$^k$University of Edinburgh, Edinburgh EH9 3JZ, United Kingdom, 
$^l$University of Fukui, Fukui City, Fukui Prefecture, Japan 910-0017
$^m$Kinki University, Higashi-Osaka City, Japan 577-8502
$^n$Universidad Iberoamericana, Mexico D.F., Mexico,
$^o$University of Iowa, Iowa City, IA  52242,
$^p$Kansas State University, Manhattan, KS 66506
$^q$Queen Mary, University of London, London, E1 4NS, England,
$^r$University of Manchester, Manchester M13 9PL, England,
$^s$Muons, Inc., Batavia, IL 60510, 
$^t$Nagasaki Institute of Applied Science, Nagasaki, Japan, 
$^u$University of Notre Dame, Notre Dame, IN 46556,
$^v$University de Oviedo, E-33007 Oviedo, Spain, 
$^w$Texas Tech University, Lubbock, TX  79609, 
$^x$IFIC(CSIC-Universitat de Valencia), 56071 Valencia, Spain,
$^y$Universidad Tecnica Federico Santa Maria, 110v Valparaiso, Chile,
$^z$University of Virginia, Charlottesville, VA  22906
$^{aa}$Bergische Universit\"at Wuppertal, 42097 Wuppertal, Germany,
$^{bb}$Yarmouk University, Irbid 211-63, Jordan
$^{jj}$On leave from J.~Stefan Institute, Ljubljana, Slovenia, 
}}
\noaffiliation